# More Secure Software Development via Cultural Analysis: How could Log4Shell happen?

**Lutz Prechelt, Freie Universität Berlin, Berlin, Germany**

**Lasse Fischer, Freie Universität Berlin, Berlin, Germany**

***Abstract*—There is little progress avoiding vulnerabilities. *Cultural analysis* may help: For a given team, find the specific cultural values that led to a past vulnerability, then modify the values accordingly for the future. After enough such studies, patterns will likely emerge that are useful for the development community at large.**

In *Anna Karenina*, Leo Tolstoy observes "Happy families are all alike; every unhappy family is unhappy in its own way." We should make use of this insight in security research. Conventional wisdom regarding how to develop software without vulnerabilities (the "happy family" case) suggests to use a standardized secure software development process. In practical software development, things go wrong from time to time and a vulnerability gets created – the "unhappy family" case. Tolstoy's insight suggests considering each of these to be unique.

At the product level (the vulnerability itself) they certainly are, but this is not likely helpful for improving the engineering process. At the process level however (how the vulnerability gets created), they may be unique in a manner that can be exploited for improving a team's engineering process and possibly also our joint secure software engineering knowledge as a community. The present article aims at explaining how.

## Vulnerability types

Let us discriminate three types of vulnerabilities by how they get created:

**Type B** ("Bug") are vulnerabilities that are due to a small, very local, everyday programming mistake.

**Type E** ("Emergent") vulnerabilities are due to an overlooked emergent property. The software was built as designed and intended, but its resulting behavior includes something that breaks security but that nobody had thought of.

**Type S** ("Slippery slope") vulnerabilities are programming mistakes, like type B. But unlike type B, this mistake is not the everyday, difficult-to-avoid minor logic error; rather, it is a mistake that is made much more likely by some aspect of the software design or the development process.

## The cultural analysis hypothesis

The present article conjectures that **many future vulnerabilities of types E and S can be avoided by studying past vulnerabilities via cultural analysis.** The approach does not help for type B. See further down for a discussion.

What is cultural analysis? Safety engineering suggests to study accidents at three levels: the **mechanics** of the accident, the **conditions** that allowed for the accident, and the **root causes** that allowed for the conditions. Root causes are very often cultural values of the development organization. **Culture** for our purposes is the set of "community-specific ideas about what is true, good, beautiful, and efficient."[2] **Cultural analysis** means uncovering the *specific* cultural values that were at work when creating a *specific* vulnerability and then reflecting how these values need to be modified for avoiding future vulnerabilities. Cultural analysis cannot primarily rely on interviews, because "why did you…"-questions are prone to produce post-hoc rationalizations that lack a true foundation. Therefore, the primary method is manually analyzing existing written materials such as code, version management history, issue tracker entries, and possibly mailing lists.

Our discussion uses the notions of **condition**, **root cause** and **culture** as backbones and hence sets them in boldface. We will now attempt to convince the reader of the truth of the hypothesis by means of a detailed example. We will return to the general level of discussion at the end of the article.

# The Log4Shell case study

The Log4Shell vulnerability, one of the biggest in software history, is of type E. It is fascinating because the underlying mistake appears not subtle at all. The vulnerability results directly from a software-architectural decision and should have been obvious. So how could it happen anyway?

Scientifically, our study is a unique-case, revelatory single-case case study in the sense of Yin[1] that is driven by a series of hypotheses, which can be considered a form of theory triangulation. The discussion of each hypothesis will draw on several pieces of evidence: data triangulation. Our account will be a simplified, linearized one. In reality, we encountered some of the material earlier and the sensemaking occurred in parallel and with a lot of confusion.

# What is Log4j?

Apache Log4j is an open source library for logging, written in Java. It is presumably the most popular of several Java logging libraries, available since 2001, and is used in very many closed-source web applications run by enterprises, among others. It is huge, about 480,000 lines of code, and has an enormous functionality regarding which types of log message originating from which places will be logged under which circumstances into which logging files and channels using which logging format. It supports many such formats and channels.

A major rearchitecting starting in 2009 turned Log4j 1.x into Log4j 2, which attempts to be more modular, crash-proof (never losing log data), and scalable. Log4j 2 should be considered a separate project, distinct from Log4j 1, and all subsequent statements refer only to Log4j 2. See Figure 1 for an overall timeline of the events described below.

# What is Log4Shell?

## Attack procedure

To exploit Log4Shell (CVE-2021-44228, disclosed to Apache 2021-11-24, CVE assigned 2021-12-05), an attacker would need to do the following:

- Set up an LDAP server publicly on the internet that serves the intended malicious code as Java bytecode.
- Pick an application to be attacked, correctly guess that it uses Log4j, correctly guess a data field that will be logged and that accepts an unrestricted string; a popular choice was the User-Agent HTTP header field that would normally identify the user's web browser.
- Send an HTTP request with a specially crafted User-Agent header value looking like this: ${jndi:ldap://attackerserver.com/maliciouscode.class}

That is all.

## Attack effect

The Log4j library will detect the ${…} syntax and perform a so-called *Lookup* on it. Lookups exist (in several flavors) to make the Log4j configuration more flexible, e.g. by

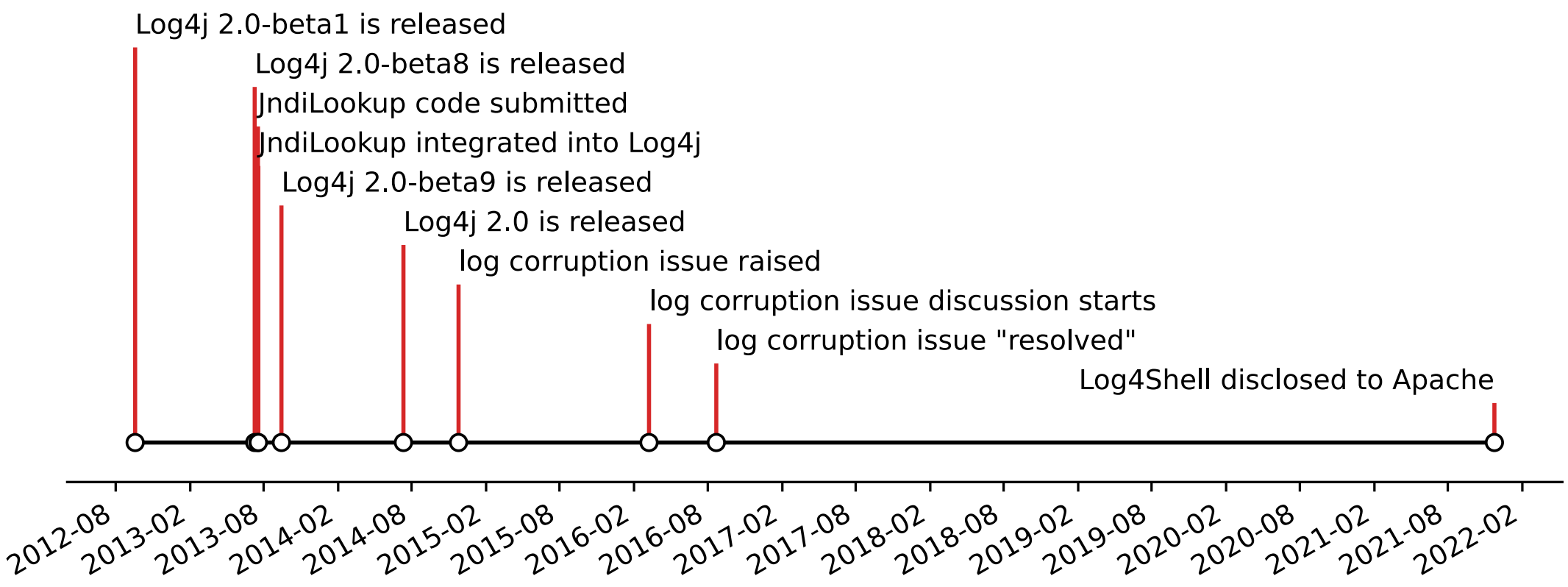

Figure 1: Timeline of events around Log4Shell insertion and discovery

forming log file pathnames derived from environment variables (*EnvironmentLookup*) or in an internationalized manner (*ResourceBundleLookup*). However, the Lookup mechanism is also typically active for the contents of log records.[4]

During a Log4Shell attack, Log4j will detect the "jndi:" prefix and activate the class *JndiLookup*. That class will supply the URL to the JNDI API of the Java standard library, which will contact attackerserver.com via the LDAP protocol, retrieve the malicouscode.class Java bytecode file and execute it.[3]

**Viability of the attack**

The attack will not always work. In particular, if the attacked server is operating in a demilitarized-zone (DMZ), a firewall will prohibit the server from contacting any outside server, including attackerserver.com.

However, many enterprise applications do apparently not operate in a DMZ.

## How could Log4Shell happen?

For our investigation we formulate and investigate (one after another) six hypotheses. We discard numbers 1 to 3, find some value in number 4, and identify numbers 5 and 6 as the most convincing explanations. When we use present tense, this will often refer to the time before Log4Shell was uncovered and repaired.

## Hypothesis 1: Complete lack of security awareness

When we started investigating how Log4Shell could happen, we expected to see an attitude towards security somewhat like the following: *"We are only a logging framework. Our functionality not likely has security implications, except perhaps for denial-of-service attacks. Our implementation language Java is memory-safe, so programming errors are also not likely to produce vulnerabilities. Security is a problem we do not have."* Such an attitude would create favorable **conditions** for the insertion of Log4Shell.

However, our investigation found that even before Log4Shell was found, the Log4j project had a security web page documenting two previous vulnerabilities with CVE identifiers in good detail and providing a security email contact point.

Hypothesis 1 is not a convincing explanation, so we dropped it.

## Hypothesis 2: Code review glitch

We next expected something like this: When somebody submitted the *JndiLookup* class for inclusion in Log4j, the code was scrutinized in a regular code review, but *nobody with a decent level of knowledge was involved in the review* to recognize the danger. This, again, would create favorable **conditions** for the insertion of Log4Shell. Checking the hypothesis can start by looking at the accident **mechanics**:

*JndiLookup* was submitted as a feature request with attached code and tests by a project outsider on 2013-07-18. It was accepted and merged the next day by the most knowledgeable person available: project founder Ralph Goers. Log4j has had over a hundred contributors overall, eight with more than 50 commits, and Goers is in the top two, having added or modified over 400,000 lines of Log4j code overall, including accepted patches from outsiders when Log4j still used Subversion, not Git. He is the mastermind behind Log4j 2's software architecture, so hypothesis 2 is also not convincing.

However, Goers' reaction to the feature request and accompanying code exhibits not the slightest trace of concern regarding security. How could Goers make such a huge mistake?

## Hypothesis 3: Release vote glitch

The Log4j project also uses a second quality gate: Roughly speaking, a prepared release candidate needs three positive votes from project members to be turned into a release. Maybe *this* was the moment where an architectural review was supposed to happen? But *only people with insufficient knowledge took part in the vote*? Again, this would create favorable **conditions** for the insertion of Log4Shell.

But the answer is no: Release processes (and votes) revolve around making sure the functionality works as intended. Reviewing, if any, concerns license issues and the test suite, not possible defects, let alone architectural ones. Furthermore, project lead Goers did take part as did Gary Gregory, the other top-2 committer.

Hypothesis 3 lacks a basis: the release vote was not supposed to stop a vulnerability.

## Hypothesis 4: Getting lost in software complexity

One thing that becomes obvious quickly to anybody involving themselves with Log4j is the high overall complexity of the software. Did maybe all participants get sufficiently lost in that complexity that they were unable to understand the danger coming with *JndiLookup*? To evaluate this hypothesis, we need to understand the software architecture.

### Which Lookups can be dangerous?

Where and when do Lookups occur? Where: Log4j Lookup expressions can occur in configuration files or in Java source code files. When: Lookup operations can be executed at loading/initialization time or at logging time. Lookups at loading/initialization time will all be defined by developers, whom Log4j needs to trust. For Lookups at logging time, the same is true if the Lookup expression stems from Java code. Lookup expressions that stem from user inputs, however, are dangerous, at least if the expression can have side effects like *JndiLookup* does.

### Log4j software architecture

For the Log4j situation in this respect, consult Figure 2.

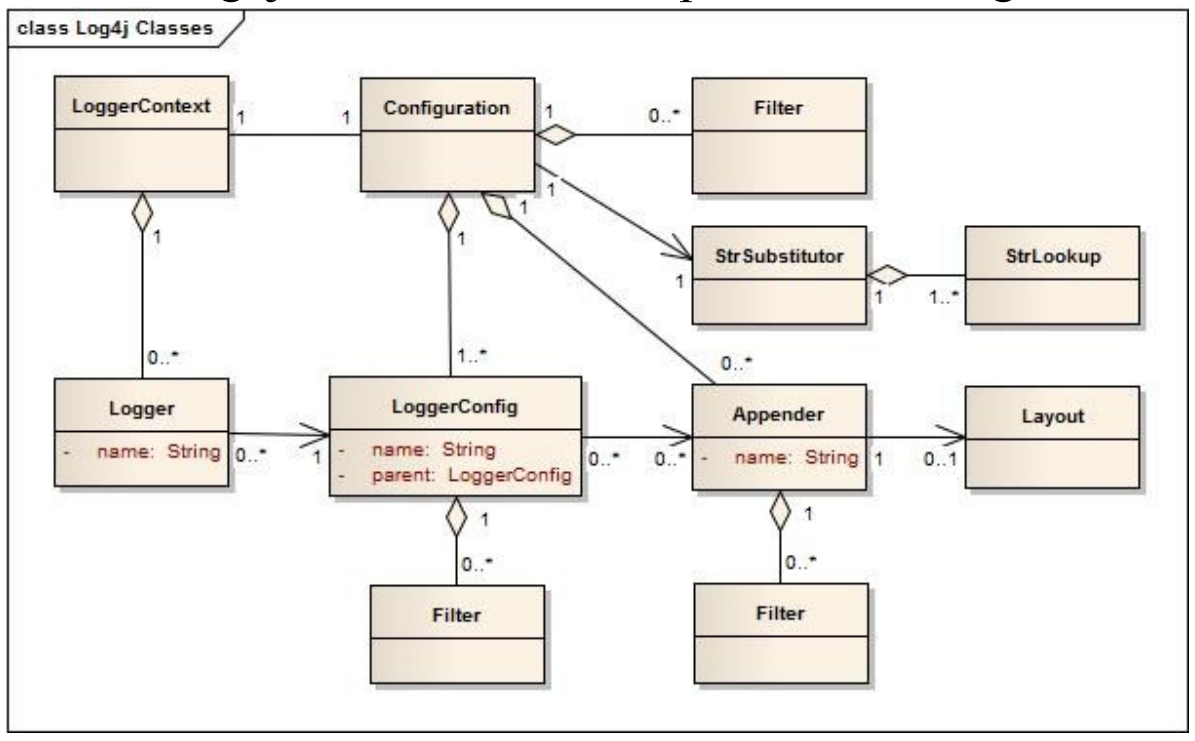

Figure 2: (taken from https://logging.apache.org/log4j/2.x/manual/architecture.html in 2023) Log4j core architecture. *Configurations* define *Loggers*. Configuration may use *Lookups* of type *StrLookup* (for instance *EnvironmentLookup* or *JndiLookup*) for specification flexibility. *Loggers* use one or more *Appenders* defining a data sink (e.g. a file or a remote syslog daemon). An *Appender* will typically use a *Layout* to format and augment the log message (e.g. turn log message "hello" into log record "2023-06-30 12:34:56,789 INFO – hello"). *Filters* can be used in several places to suppress the logging of certain records.

The interesting element of this is *Layout*. Its by far most commonly used implementation is *PatternLayout*, which allows specifying the format "2023-06-30T12:34:56.789 INFO – hello" as "%d %level – %m". There are many more such conversion patterns.

The reason why all this is interesting is that the figure does not mention one very important fact: Not only *Configuration* will apply any *StrLookup* (including *JndiLookup*), *PatternLayout* will do this as well.[4]

### Log4j documentation

From a puristical point of view, this lack of mention is justifiable: Lookups do not occur in *Layouts* in general, only in *PatternLayout* – which is not shown in the Figure because Log4j has about a dozen *Layout* classes, most of them highly specialized. From a pragmatic point of view, however, *PatternLayout* and its behavior are so important that the documentation gap is deplorable.

But it gets worse: Mentioning *Lookups* in the context of *Layout* is missing not only from Figure 2. It is also missing in most of the textual documentation of *Lookups*, which starts with the sentence *"Lookups provide a way to add values to the Log4j configuration at arbitrary places."*. To the configuration, not the log records! Across the voluminous Log4j documentation, only a single sentence in a semi-obscure paragraph about the provenance of some components ever mentions modifying log records. A vast majority of Log4j users will remain unaware that *Lookups* may occur in log records.

### Evaluation

If you are like us, you will find all this complicated and confusing. Will it be less confusing for Log4j developers? Sure, they spend more time with it. But they are also confronted with many more parts than we: Besides the above-mentioned dozen Layout classes, Log4j has about 18 Filter classes, about 30 Appender classes, and lots and lots of internal infrastructure not mentioned above at all.

For instance, the implementation of *PatternLayout* alone contains 7 internal helper classes, imports about 18 classes from elsewhere, and has over 700 lines of code. Keeping in one's head, even approximately, what all these parts are doing, would be a considerable feat. These are favorable **conditions** for something like Log4Shell to happen.

### Additional evidence

In 2018-01, the team discussed what to do about the increasing size of the Log4j codebase[11]. A team member listed the modules from the Java standard library that the log4j-core module was already using. One entry said "java.naming: JNDI", to which Ralph Goers replied *"Where are we using JNDI?"*. He had forgotten about the functionality he had incorporated himself. The question received the correct answer *JNDILookup*, but nobody noticed any problem.

### So what?

Hypothesis 4 clearly has credibility: Log4j does have a complexity problem. However, the above analysis is rather unspecific; we do not learn much from it. Therefore, we keep searching.

## Hypothesis 5: Log4j is suffering from featuritis

### How to recognize the vulnerability

For recognizing JndiLookup as a vulnerability, developers must know and be aware of two crucial facts:
**F1**: A 'jndi:' lookup expression can not only result in a string replacement (if it returns a string); it can also have the side-effect of loading remote code and executing it[3] (if it returns a Java class).
**F2**: Lookups are active not only in configuration, but also in PatternLayout and hence usually for entire log records.[4]
Not knowing or overlooking either of these two facts creates a **condition** that makes it hard to recognize the vulnerability. Has this happened? Was it not knowing or was it overlooking? And of which of the two facts?
The following development episode sheds light on this.

### The log corruption episode

2014-11 (16 months after the creation of Log4Shell), Moritz Löser, an outsider to the Log4j project, created an issue[5] in the bug tracker titled *"Ability to disable (date) lookup completely, compatibility issues with other libraries like Camel"*. He complained that, when he had tried *"to log stuff like this:* `log.info("${date:now:buhu}")` *[...] log4j2 tries to change the logged output - want to substitute a date. This is intransparently limiting the logging.*
*nobody knows what kind of strings are allowed and what kind of strings will be interpreted by log4j2. [...] The only way i see is a switch to disable all kinds of lookups and it should probably [be] off by default."*
Note how this posting gets several things right[4]: (1) The lookup functionality for log records is not documented. (2) It should be an optional functionality. (3) Its default should be "off". With an "off" default, Moritz Löser would be happy and Log4Shell would not exist.
It takes 16 months until the discussion of this issue starts in 2016-03, but then all four top-4 Log4j contributors participate, so at least from now on the project is most certainly aware of fact F2. One of the four, Gary Gregory, comments in 2016-08 *"If we provide/document a way to disable lookups, then we should probably make the distinction between lookups being used in configuration files and in messages."*, a good addition to the three points above, let us call it (4). Ralph Goers, however, makes a different suggestion: extending the %m placeholder of *PatternLayout* to allow writing %m{nolookups}, which would suppress *Lookups* for only the log records created by this particular log statement. Gary Gregory agrees and implements that solution the next day. End of episode.

### Interpretation

Apache projects have a meritocratic-democratic culture. They use an approach called "lazy consensus[6]" by which sensible ideas get accepted, except that better ideas are preferred when they get proposed, and bad ideas are usually superseded by sensible ones. Log4j works like this, too. So the reason why Ralph Goers' proposal gets accepted is surely *not* the fact that he is the project founder, the reason must be something else. But what? We believe (and offer evidence below) that the Log4j project community has a strong **cultural** value stating that more functionality is better than less functionality. With such an attitude, the devastating "on" default for *Lookups* in *PatternLayout* may feel natural and even preferable.

### Evidence for the "more functionality is better" attitude

We offer six rather diverse pieces of evidence.
First, let us start with the above episode: Points (2) to (4) are normative, they recommend (2) making Lookups optional, (3) having them default to off, and (4) doing this separately for config and for messages. The solution that subsequently got implemented follows (2) and (4), but not (3), which would have avoided the Log4Shell disaster five years later but also would have deactivated existing functionality.
Second, when Log4j finally did deactivate the lookup functionality for log messages, the issue[7] proposing it offered three justifications: (a) it is confusing, (b) it differs from other implementations of java.util.logging, (c) it costs execution time. Of these, (a) should have been the main one. In the issue tracker entry, however, it only consists of the single word "confusing", while the other two are elaborated upon with a total of 109 words.
Third, even that change[7] did only *deactivate* the functionality, not remove it. It could still be turned on with a configuration flag.
Fourth, it is telling why the original author of Log4j 1 never entered the Log4j 2 project. In an interview[8] he explained that he used to reject 95 percent of all change requests, while Ralph Goers had a more "liberal attitude" in that respect.
Fifth, in an unrelated and not atypical episode[9] in 2013-05, an outsider requested to make an includeLocation() method protected (not private) so they could use it when overriding LoggerConfig. Goers replied that overriding LoggerConfig was not recommended and suggested a more complex alternative mechanism for achieving the requesters goal. Ten days later he reported to have implemented *both*.

Finally, in another unrelated episode[10] in 2014-04, a new project member suggested that log record markers should be mutable so they could have multiple parents. Goers replied that they were deliberately immutable, because mutability could create huge performance issues. A complex development episode ensued, involving several design ideas and multiple rounds of prototyping and benchmarking. It finished with a Marker implementation that could indeed represent multiple parents and created only little performance loss, but was also 300 lines long.

It is probably fair to conclude from all this that Log4j's urge for extending its functionality is strong – so strong, that when a feature like *JndiLookup* is offered, it is almost irresistible unless it is recognized as deeply problematic. In that case, the urge for functionality could be considered a **root cause**. But *JndiLookup* of course *should* have been recognized as deeply problematic, or not?

## Hypothesis 6: Log4j is suffering from Java's featuritis as well

As annoying as the log corruption effect likely has been for Moritz Löser: By itself, F2 is relatively harmless. The dangerous part comes mostly from F1 – Lookups should not have side effects.

F1, however, is a relatively obscure fact: Few people made active use of class loading via JNDI queries when Log4Shell was inserted (nor today) and there were hardly any hints anywhere in the JNDI-related API documentation or tutorials that JNDI queries may execute remote code, let alone an explicit discussion of how to prevent it.

Knowing F1 required a JNDI expert; it should not be a surprise that no one in the Log4j project recognized the problem.

Looking at this issue from the Java side of things shows a similar picture: The ability to load code via JNDI and execute it is clearly a cool feature, but just as clearly it should be turned off by default. Why does it default to on?

The history of Java suggests the following **cultural** value as the answer: The JNDI remote code loading capability defaults to on because Java library classes tend to be as powerful as anybody involved was able to imagine them.

In our case: JNDI queries can return different types of data? Then they should be able to return *any* type of data! And that includes objects of type Class. Java works like this in countless places, the second **root cause**.

Specifically, the LDAP implementation of Java SE 7, the version that was current when Log4Shell was introduced, had no switch at all for turning off remote code loading.

During the lifetime of Log4Shell, three long-time support (LTS) versions of Java SE appeared that are all still in wide use today: Java SE 8, 11, and 17. Version 8 introduced the global flag com.sun.jndi.ldap.object.trustSerialData, but, in an uncanny emulation of the Log4j project's behavior, its default was *true*. It was still *true* in version 11. It was still *true* in version 17.

Only the current version, Java SE 20, which appeared *after* Log4Shell was discovered, finally uses a default of *false*.

So we might say that Log4Shell was not entirely the Log4j project's fault: Java and Oracle also contributed a lot. This means the second **root cause** lies outside the Log4j project itself.

## Log4Shell case study summary

### Non-reasons for Log4Shell

In a nutshell, our investigation found that Log4Shell was not created due to a general lack of security awareness in the Log4j project (hypothesis 1); such awareness exists. It was also not produced due to a glitch in the quality assurance processes (hypotheses 2 and 3); these worked normally.

The enormous size of the Log4j code base (hypothesis 4), appears to be problematic, but not a key factor either; it has contributed to favorable **conditions**, but is not a **root cause**.

### Reasons for Log4Shell

Instead, we found two **root causes**:

1. Log4J appears to have a strong **cultural preference that more functionality is better than less functionality** (hypothesis 5). This preference appears to have produced the strange decision of enabling *Lookups* in log records by default and the even stranger decision to keep this default even after a user complained that this had ruined his output and was not even documented. The much more sensible (from a general user point of view) default of "off" would have prevented Log4Shell.
2. The Java SE platform itself appears to have a **similar culture** and behaviors (hypothesis 6).

There, these preferences first produced overly powerful JNDI functionality that was not only able of retrieving and executing remote code, but would even *default* to doing so whenever the directory server returned a code object. Later on, they resulted in a retro-fitted security mechanism with an insecure default setting that eventually converged to a secure default only a dozen years later. Too late for preventing Log4Shell.

## Log4Shell case study conclusions

We hope it is now obvious that for the specific case of the Log4Shell vulnerability, cultural analysis is likely a helpful exercise. It uncovers a root cause that the Log4j team can then keep in check through future reflection when they make decisions. Completely removing the root cause is likely unrealistic, but is not needed.

For instance, when the team reads about the log corruption episode from the root cause perspective, they can easily see how they should extend and shift their priorities to make it less likely that another vulnerability arises from their (continued) “more functionality is better” inclination.

And when the team understands Java’s “more powerful is better” cultural value, it can develop a habit of approaching the use of any never-before-used feature of Java with caution (which admittedly would probably not have helped in the specific case of JNDI; Java should get half of the blame for Log4Shell).

## Validation by member check

We asked the top-4 committers of the Log4j team to read a draft of this article and report whether a) they agree with the root cause analysis and b) they learned anything from it that will help them avoid future vulnerabilities.

Two of them responded: Ralph Goers and Matt Sicker. They both confirmed the facts and root causes we report. They did not report specific learnings but both of them stated that they believed cultural analysis was a helpful technique; Sicker called it “very powerful”.

## So can cultural analysis help?

Now let us switch the perspective away from the specific case of Log4Shell to the general case of any vulnerability anywhere. What can cultural analysis contribute for teams with less spectacular, more mundane vulnerabilities?

### Looking at the CWE Top 25

Reviewing the Common Weakness Enumeration (CWE) list of the most frequent weaknesses that lead to vulnerabilities does not suggest at first that many vulnerabilities can be avoided by cultural analysis of past ones. Rather, it reads like a list of important gaps in software engineers’ education: Most vulnerabilities arise from only a handful of different types of software behavior that are all well-known und well-understood.

One may get the impression that software engineers simply need to learn about these and then pay attention to them and then most vulnerabilities would be avoided.

### CWE is not about development process

Upon second look, however, this impression is incorrect. The CWE classes consider only the software and its behavior, not the process how that software came to be. For instance, the distinction made in our simple discrimination of vulnerability types B, E, and T above is entirely missing from CWE.

Indeed, even the spectacular Log4Shell case looks banal through the lens of its CWE classes: Apache found it to be Uncontrolled Resource Consumption (CWE-400), Deserialization of Untrusted Data (CWE-502) and Improper Input Validation (CWE-20).

In contrast, the process perspective of cultural analysis adds a lot more detail: We recognize that the problem with deserialization was to not *know* how powerful the underlying Java apparatus is. The problem with input validation was to have *forgotten* that a Lookup could happen in untrusted log records not only in trusted configuration.

### For type B (simple “bug”) vulnerabilities

As stated above, cultural analysis is not likely helpful for avoiding small, isolated programming mistakes. Learning about what are the most dangerous types of these mistakes and then paying attention to them will continue to be the only practical path leading away from type B vulnerabilities for most teams.

### For type E (“emergent”) vulnerabilities

Log4Shell is a type E vulnerability and cultural analysis appears helpful there. We conjecture this will be so for most type E cases, because even emergent properties are knowable approximately. Therefore, to be unpleasantly surprised by one likely involves failing to perform this approximate looking-ahead.

### For type S (“slippery slope”) vulnerabilities

We conjecture that slippery slope effects will be a large and interesting set of situations to dissect via cultural analysis and hope such an analysis will often motivate the team to work against the slope in order to get rid of it or at least reduce it. If so, cultural analysis ought to be useful for this class as well. In fact, Log4Shell could also be viewed as a type S problem, because indiscriminately accumulating more and more functionality is a known type of slippery slope.

### Team-specific vs. general insights: The CCE

So far, we have only argued that cultural analysis can help specific teams to get better in their development practices. But will it help the larger software development community as well? We expect the following to happen: Once a few dozen cultural analyses have been performed, patterns start to emerge in the cultural values underlying the vulnerabilities. Once these become clear, they could be transformed into a CCE (Common Cultural-Value Enumeration). Along the huge CVE and the large CWE, we expect the CCE to be small, with only a few dozen entries. A CCE would help teams perform cultural analysis proactively and as a self-assessment and so might help prevent vulnerabilities more broadly.

## Conclusions and next steps

### Take-home messages

First, nailing down the *specific* cultural factors (and their interplay) that led to a vulnerability, instead of only mumbling "security culture" in general terms, allows understanding how (in the sense of root causes, not mere mechanisms) the vulnerability was created. This understanding is constructive in the sense that it allows the respective team to recognize dangerous habits and tendencies in order to unlearn them and move to safer ones.
Second, it should not be difficult to boil a list of such analyses down to a general list of typical specific cultural problems in development teams that lead to vulnerabilities, the Common Cultural-Value Enumeration (CCE). The CCE can be used in education and training at the process level much like the CWE can be used at the product level.

### What is next?

For working towards a CCE, we need to perform many more cultural analyses and need to understand better how to perform them and make use of them.
*Should a cultural analysis be performed entirely from the outside, as we did here, or is it better to obtain some information directly from team members?* The latter could save a lot of effort, but also might misdirect the researchers' attention. For instance in a conversation with Ralph Goers during the member check, he said that in his view, the "more functionality is better" tendency was a secondary effect, not a primary one. The underlying primary effect was his focus on having a large and vibrant development community in Log4j. He expected that such a community would unavoidably produce lots of new code and additional functionality. Such information may be helpful, but may also be distracting.
*Will teams be willing and able to pick up outcomes of a cultural analysis and move on with their culture?* We expect this to be a difficult problem initially, until the idea becomes mainstream. For example, the Log4j team never asked us for ideas how they might do that. Rather, Ralph Goers told us he could not remember why he had included code that would perform Lookup expansion on log records. He also said that "nobody in their right minds" would expect a JNDI query to download code and execute it, although that behavior is a natural consequence of a fully general Java serialization-deserialization facility when applied remotely. Both of these details do not suggest much eagerness to learn about the Log4j team's culture. Work will be needed to learn how to break this up or how to identify interested teams upfront; involving the team in the analysis will likely be helpful in this regard.
Answering these questions can go hand-in-hand with the work on accumulating more cultural analyses. For that, we need to identify a list of promising vulnerabilities for which to perform a cultural analysis. Helpful properties will be: not being too old, enough detail available about the development process (e.g. many lively discussions in the issue tracker, whether open source or proprietary), and ideally a development team interested in learning more about its way of working.

## REFERENCES

1. R.K. Yin, "Case study research: Design and methods", Sage, 2009.
2. R.A. Shweder and L. Beldo, "Culture: contemporary views", in: International Encyclopedia of the Social & Behavioral Sciences, 2nd edition, Volume 5, pp. 582-589, 2015.
3. A. Muñoz and O. Mirosh, "A Journey From JNDI/LDAP Manipulation to Remote Code Execution Dream Land", Presentation at Blackhat USA, https://www.blackhat.com/docs/us-16/materials/us-16-Munoz-A-Journey-From-JNDI-LDAP-Manipulation-To-RCE.pdf, 2016.
4. J. M. Edwards: "Log4j vulnerability: What the FAQ happened?", https://jedwidz.hashnode.dev/log4j-vulnerability-what-the-faq-happened, Dec 15, 2021.
5. M. Löser, M. Sicker, R. Goers, M. Gysel, L. Burgazzoli, G. Gregory, R. Popma, "Ability to disable (date) lookup completely, compatibility issues with other libraries like Camel", Log4j issue tracker,

https://issues.apache.org/jira/browse/LOG4J2-905, Nov 2014 to Aug 2016.
6. Apache Software Foundation, "How the ASF works", Section "Decision making", https://apache.org/foundation/how-it-works/#decision-making, 2004.
7. C. Kozak, "Message lookups should be disabled by default", Log4j issue tracker, https://issues.apache.org/jira/browse/LOG4J2-3198, Nov 2021.
8. R. Fulterer: "Log4j wurde 1997 in der Schweiz entwickelt – der Erfinder erzählt" (Log4j was developed in Switzerland in 1997 – the inventor recounts), Neue Zürcher Zeitung, https://www.nzz.ch/technologie/log4j-wurde-1997-in-der-schweiz-entwickelt-der-erfinder-erzaehlt-ld.1660571, Dec 16, 2021.
9. K. Lehrke, G.D. Gregory, R. Goers, "LoggerConfig has a method that should be protected instead of private", Log4j issue tracker, https://issues.apache.org/jira/projects/LOG4J2/issues/LOG4J2-243?orderby=created+ASC, May 2013.
10. B. Brouwer, R. Goers, G.D. Gregory, "Markers not as powerful as slf4j", Log4j issue tracker, https://issues.apache.org/jira/projects/LOG4J2/issues/LOG4J2-585?orderby=created+ASC, Mar/Apr 2014.
11. The Log4j development team, "The shape of Log4j", Log4j mailing list, https://lists.apache.org/thread/npld5rc272g6ozmonqx8m3d1vzv3jc4f, Jan 2018.

**Lutz Prechelt** is professor for software engineering at Freie Universität Berlin. His research interests concern software development processes, in particular the human factor, and research quality. He is the founder of the Software Engineering Research Quality Coalition (SERQco). Contact: prechelt@inf.fu-berlin.de.

**Lasse Fischer** received a Bachelor in Informatics from Freie Universität Berlin. He performed parts of the investigation described above as his Bachelor thesis. Contact: lasse.fischer@gmx.net.